\documentclass[10pt,conference]{IEEEtran}
\usepackage{graphicx} 
\usepackage{amsmath}
\usepackage{amssymb}
\usepackage[colorlinks,linkcolor=blue,pdfborder={0 0 0},citecolor=magenta,bookmarks=false,hypertexnames=true]{hyperref}
\usepackage{braket}
\usepackage[thinc]{esdiff}
\usepackage[citestyle=ieee,sorting=none,sortcites=true,doi=true,url=false,giveninits=true,hyperref,maxbibnames=3,minbibnames=1,style=ieee,abbreviate=false]{biblatex}
\usepackage{orcidlink}

\begin{document}
\title{Arbitrary Ground State Observables from Quantum Computed Moments}

\author{\IEEEauthorblockN{\orcidlink{0009-0007-3355-5458} Harish J. Vallury\IEEEauthorrefmark{1} and
\orcidlink{0000-0001-7672-6965} Lloyd C. L. Hollenberg\IEEEauthorrefmark{2}}
\IEEEauthorblockA{School of Physics,
The University of Melbourne\\
Parkville, VIC 3010, Australia\\
Email: \IEEEauthorrefmark{1}hvallury@student.unimelb.edu.au,
\IEEEauthorrefmark{2}lloydch@unimelb.edu.au
}}

\maketitle

\begin{abstract}

The determination of ground state properties of quantum systems is a fundamental problem in physics and chemistry, and is considered a key application of quantum computers. A common approach is to prepare a trial ground state on the quantum computer and measure observables such as energy, but this is often limited by hardware constraints that prevent an accurate description of the target ground state. The quantum computed moments (QCM) method has proven to be remarkably useful in estimating the ground state energy of a system by computing Hamiltonian moments with respect to a suboptimal or noisy trial state. In this paper, we extend the QCM method to estimate arbitrary ground state observables of quantum systems. We present preliminary results of using QCM to determine the ground state magnetisation and spin-spin correlations of the Heisenberg model in its various forms. Our findings validate the well-established advantage of QCM over existing methods in handling suboptimal trial states and noise, extend its applicability to the estimation of more general ground state properties, and demonstrate its practical potential for solving a wide range of problems on near-term quantum hardware.

\end{abstract}

\section{Introduction}
Many difficult problems with practical application in physics and chemistry involve determining the ground state properties of a quantum system. This has been a popular road to follow towards the development of near-term applications of quantum computers, with much of the recent focus primarily on variational quantum algorithms~\cite{cao2019quantum,mcardle2020quantum,cerezo2021variational}. These hybrid approaches involve the preparation of complicated states designed to represent the ground state of the desired quantum system, via the measurement and minimisation of energy~\cite{vqe,kandala,nam2020ground}. Most interesting problems at scale require high-depth circuits to prepare such states, and are thus subject to noise which is often severely detrimental to the quality of results~\cite{wang,stilck2021limitations}.

Some time ago, an analytic cluster expansion of the Lanczos recursion was uncovered~\cite{pexp}. Following from this, it was found that for a given quantum system Hamiltonian $H$, one could compute Hamiltonian moments $\langle H^k \rangle$ with respect to a trial state with some overlap with the ground state to produce a powerful approximation to the ground state energy \cite{el4,analytic}, shown below to fourth order in the moments:
\begin{equation}
E_{\rm 0}^{\rm{L}(4)} = c_1 - \frac{c_2^2}{c_3^2 - c_2 c_4} \left[\sqrt{3 c_3^2 - 2 c_2 c_4} - c_3\right], \label{formula}
\end{equation}
where the cumulants $c_n$ are defined as:
\begin{equation}
c_n = \langle H^n\rangle - \sum_{p=0}^{n-2}\binom{n-1}{p}c_{p+1} \langle H^{n-1-p}\rangle.
\end{equation}

The rapid development of quantum computer technology has provided a fitting application for this result, allowing for advanced state preparation and measurement of these Hamiltonian moments in order to solve interesting problems. We refer to this as the quantum computed moments (QCM) method, and recent work applied to problems in quantum magnetism~\cite{hv2020, Vallury2023noiserobustground} and chemistry~\cite{mj-chem} has demonstrated the effectiveness of QCM in estimating the ground state energy of a system, even when moments are measured with respect to a suboptimal or noisy trial state. This technique can correct for a low complexity ansatz and a high degree of noise, proving to be remarkably useful on present-day quantum hardware for a variety of different problem instances.

In recent times, these QCM studies have been accompanied elsewhere in the literature by a general renewed interest in Hamiltonian moments in a quantum computing context \cite{kowalski2020quantum,seki2021quantum,suchsland2021algorithmic,momentsreview}, with the primary focus being on the ground state energy problem. However, a problem which is much more practically useful and indeed more computationally complex is that of determining arbitrary ground state observables~\cite{zhang2022computing}. It turns out that Hamiltonian moments can also be used to estimate general ground state properties via a Hellman-Feynman approach~\cite{hollenberg1994staggered,witte1997two}. Here we develop this technique in the QCM context. When applied to models of interest, our results show that the crucial robustness property of QCM for energy problems indeed carries over to this more general ground state observable procedure.

\section{Method}
Suppose we wish to find the expectation value of some general observable (given by Hermitian operator $A$) with respect to the ground state $\ket{\psi_0}$ of a Hamiltonian $H$. Consider the following parameterised Hamiltonian:
\begin{align}
    H_{\lambda} = H + \lambda A \label{eq1}
\end{align}

The parameterised Hamiltonian $H_{\lambda}$ has ground state $\ket{\psi_{\lambda,0}}$ with energy $E_{\lambda,0}$ which we can approximate with:
\begin{equation}
    E_{\lambda,0} \approx E^{{\rm L}(4)}_{\lambda,0}|_{\psi_t}.\label{eq2}
\end{equation}
The approximation in Equation (\ref{eq2}) is expressed entirely in terms of Hamiltonian moments $\langle \psi_t |(H_{\lambda})^k| \psi_t \rangle$, up to $k=4$, with respect to some trial state $\ket{\psi_t}$ which has nonzero overlap with $\ket{\psi_{\lambda,0}}$.

Starting with the Hellmann-Feynman theorem,
\begin{equation}
     \langle \psi_{\lambda,0}|\,\diff{H_\lambda}{\lambda}\,|\psi_{\lambda,0} \rangle = \diff{E_{\lambda,0}}{\lambda},
\end{equation}
one can substitute the derivative of the Hamiltonian in Equation (\ref{eq1}) and use the approximation in Equation (\ref{eq2}) to obtain:
\begin{equation}
     \langle \psi_{\lambda,0}|\,A\,|\psi_{\lambda,0} \rangle \approx \frac{\rm d}{\mathrm{d}\lambda}\left(E^{{\rm L}(4)}_{\lambda,0}|_{\psi_t}\right). \label{eq4}
\end{equation}
Thus if we want to compute the observable $A$ with respect to the ground state $\ket{\psi_0}$ of $H$, we can choose $\ket{\psi_t}$ to have some overlap with $\ket{\psi_0}$ and evaluate the derivative in Equation (\ref{eq4}) around $\lambda=0$:
\begin{equation}
    \langle \psi_{0}|\,A\,|\psi_{0} \rangle \approx \frac{\rm d}{\mathrm{d}\lambda}\left(E^{{\rm L}(4)}_{\lambda,0}|_{\psi_t}\right)_{\lambda=0}. \label{eq5}
\end{equation}
This leads to a practically useful final expression for the ground state observable:
\begin{equation}\label{AL}
    \langle A \rangle_0 ^{\rm L(4)} \equiv \frac{1}{2\varepsilon}\left(E_{+\varepsilon,0}^{\rm L(4)}-{E_{-\varepsilon,0}^{\rm L(4)}}\right) \approx \langle \psi_{0}|\,A\,|\psi_{0} \rangle,
\end{equation}
where $\varepsilon>0$ is sufficiently small as to find the $\lambda=0$ derivative, and the $E_{\pm \varepsilon,0}^{\rm L(4)}$ are calculated via Equation (\ref{formula}) from the moments $\langle \, (H \pm \varepsilon A)^k \, \rangle$ with respect to some trial state $\ket{\psi_t}$ up to $k=4$.

We expand upon the analytic studies of this method, first introduced in~\cite{hollenberg1994staggered,witte1997two}, by now considering the approach in a quantum computing context. Note that, in practice, the $E_{+\varepsilon,0}^{\rm L(4)}$ and ${E_{-\varepsilon,0}^{\rm L(4)}}$ in Equation~\ref{AL} are not computed separately, but rather come from the same set of measurements on the quantum computer. The derivative is taken with respect to a parameter only introduced in the classical post-processing stage, so remains unaffected by the stochastic nature of the quantum measurement. We can thus expect to see similar robustness behaviour as has been observed in the QCM energy estimate, and in the following section, we test the effectiveness of this arbitrary observables technique by performing noisy simulations on a variety of trial states and Hamiltonians.

\section{Results}

\begin{figure}[ht]
\includegraphics[width=0.45\textwidth]{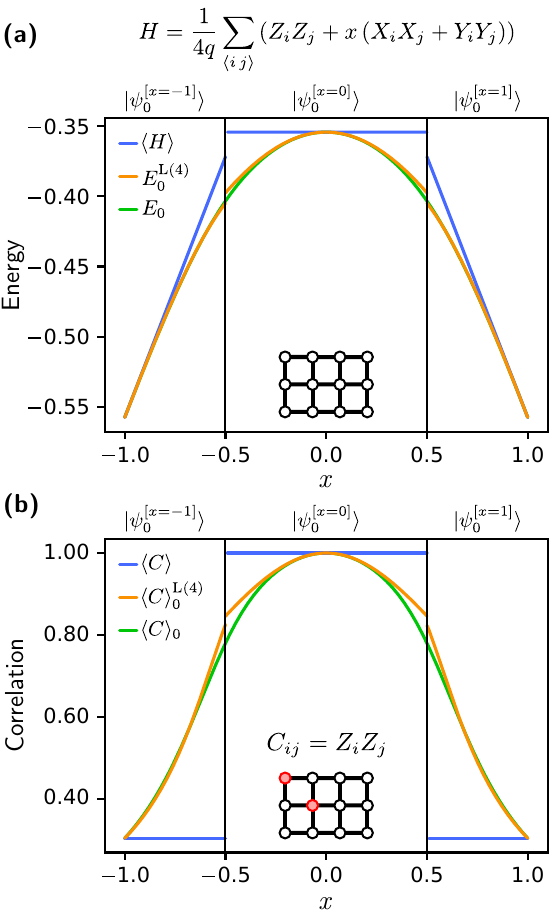}
\centering
\caption{\textbf{(a)}~Ground state energy estimates for the Heisenberg XXZ model as a function of $x$ in the Hamiltonian. Green shows the exact ground state energy, $E_0$. The energy estimates, $\langle H \rangle$ (blue) and $E_0^{\rm L(4)}$ (orange), are evaluated with respect to three trial states: the $x=-1$ (ferromagnetic) ground state for $-1<x<-\frac{1}{2}$, the $x=0$ (N\'eel) ground state for $-\frac{1}{2}<x<\frac{1}{2}$, and the $x=1$ (antiferromagnetic) ground state for $\frac{1}{2}<x<1$. \textbf{(b)}~Estimates of ground state $ZZ$ correlation between two next-nearest-neighbour spins (marked in red) versus $x$, using the same three trial states. Like before, the exact correlation is shown in green, the direct expectation value in blue, and the moments-based estimate in orange.}
\label{Fig1}
\end{figure}

\begin{figure*}[ht]
\includegraphics[width=0.98\textwidth]{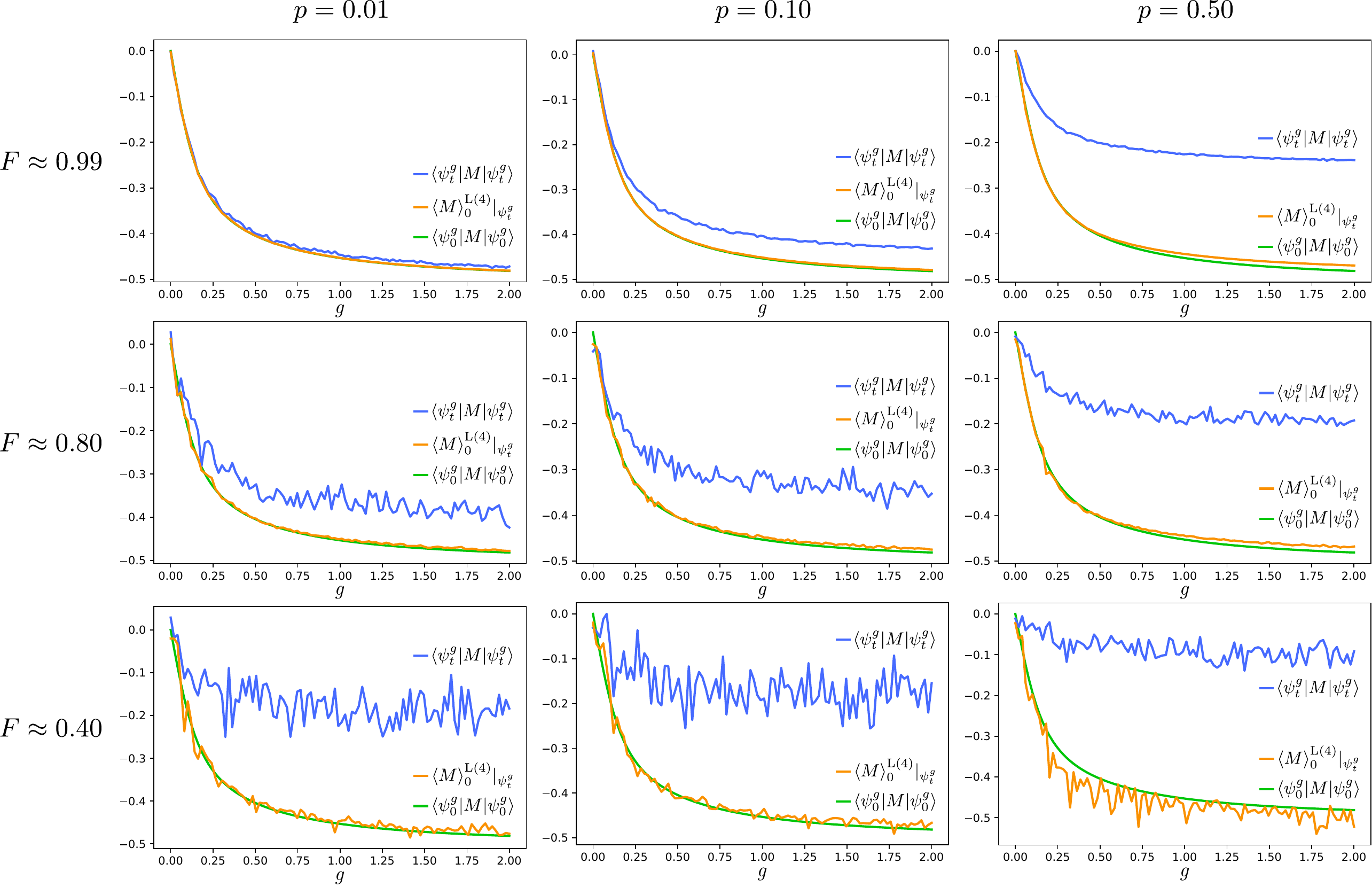}
\centering
\caption{Staggered magnetisation for a 6-site 1D Heisenberg antiferromagnet with staggered external field, as a function of field parameter $g$ in the Hamiltonian (Eq.~\ref{Hs}). Comparison of methods for obtaining ground state staggered magnetisation -- direct $\langle M \rangle$ (blue) and QCM $\langle M \rangle_0^{\rm L(4)}$ (orange) estimates vs. exact value (green). To compute the estimates, trial states $\ket{\psi_t^g}$ were generated at each value of $g$ by starting at the ground state $\ket{\psi_0^g}$ and applying a random rotation as $\ket{\psi_t^g} = e^{-i\theta M}\ket{\psi_0^g}$, where $M$ is a random Hermitian matrix sampled from the Gaussian unitary ensemble. By tuning the parameter $\theta$, trial states of varying degrees of overlap with the ground state $F = \left|\langle \psi_t^g | \psi_0^g \rangle\right|^2$ were generated. These trial states ${\rho}=\ket{\psi_t^g}\!\bra{\psi_t^g}$ were then passed through a depolarising error channel ${\rho} \mapsto (1-\frac{3p}{4}){\rho} + \frac{p}{4}(X{\rho}X+Y{\rho}Y+Z{\rho}Z)$ with noise parameter $p$, which was varied from 1\% up to 50\%.}
\label{Fig2}
\end{figure*}

Of particular interest in the study of quantum magnetism is the Heisenberg XXZ model, given by the Hamiltonian:
\begin{equation}
    H = \frac{1}{4q}\sum_{\langle i\,j\rangle}\left(Z_iZ_j + x(X_iX_j+Y_iY_j)\right).
\end{equation}
Here, $q$ is the number of qubits, the $S_i$ are the Pauli operators $S\in\{X,Y,Z\}$ acting on the $i^{\rm th}$ qubit, the $\langle i\,j\rangle$ are the nearest neighbour couplings in the lattice. The parameter $x$ controls the strength of the transverse spin components relative to the Ising-like interaction. As a first test case of the moments-based ground state observable estimate (Eq.~\ref{AL}), we take the XXZ model over a 2D square lattice, and consider the $ZZ$ correlation operator,
\begin{equation}
    C_{ij} = Z_iZ_j,
\end{equation}
acting on a pair of next-nearest-neighbour qubits $i$ and $j$. Focusing on these longer range interactions provides additional insights into the system behaviour beyond what is captured explicitly by the Hamiltonian.

On a quantum computer, to simulate and study the ground state behaviour (e.g. the energy and spin-spin correlations) of this system as the parameter $x$ varies, one might need to prepare the ground state using a different circuit at each value of $x$. However, one can perform far fewer computations by leveraging the fact that the QCM approach has reduced sensitivity to the choice of trial state. A well-chosen state that accurately describes the solution at a given value of $x$ will also produce an accurate QCM estimate for nearby values of $x$. Figure~\ref{Fig1}(a) shows that the ground state energy of the Heisenberg XXZ model can be fully characterised for all $-1<x<1$ using the QCM estimate $E_0^{\rm L(4)}$ with respect to just three different trial states (ground states at $x\in\{-1,0,1\}$), in stark contrast to $\langle H \rangle$ from the standard variational approach.

In Figure~\ref{Fig1}(b), the difference between the two methods is made even clearer when these states are used to find estimates of other ground state properties. QCM correctly estimates the correlation in the ground state between the corner site and its next-nearest-neighbour in the middle of the $4\times3$ square lattice, remaining remarkably close to the exact value as $x$ varies, diverging only slightly at $x=\pm\frac{1}{2}$ where the ground state overlap is at a minimum.

In practice on a real device, measuring the moments up to $H^4$ requires one to multiply out the terms (or Pauli strings) in the Hamiltonian and determine their expectation values. These Pauli strings can be grouped into simultaneously measurable tensor product basis (TPB) sets, the total number of which scales logarithmically with the number of Pauli strings \cite{hv2020}. Each TPB corresponds to one measurement on the quantum computer -- thus when determining a general ground state observable using QCM, if the operator describes interactions present explicitly in the Hamiltonian, no additional measurements need to be made over the energy calculation. In our Heisenberg XXZ case, to go from the energy calculation in Figure~\ref{Fig1}(a) to the next-nearest-neighbour correlation in Figure~\ref{Fig1}(b), the total number of Pauli strings increases from 66\,343 to 68\,960, and the total number of TPB measurements goes from 1\,906 to 1\,973. Note that these measurements need to only be made once for each of the three trial states, after which the entire curve for $-1<x<1$ can be filled out via simple classical post-processing.

So far, we have shown that the QCM estimate of a ground state observable other than energy shows the same resilience to an imperfect choice of trial state, however the presence of the noise robustness property still remains to be verified. For this, we turn our attention to the antiferromagnetic Heisenberg model with an external (staggered) field:
\begin{equation}\label{Hs}
    H = \frac{1}{4}\sum_{\langle i\,j\rangle}(X_iX_j + Y_iY_j + Z_iZ_j) + \frac{g}{2}\sum_k (-1)^kZ_k,
\end{equation}
where $g$ is the external field parameter. An operator of interest in this case is the staggered magnetisation $M$, given by:
\begin{equation}\label{Ms}
    M = \frac{1}{2}\sum_k (-1)^kZ_k.
\end{equation}

We look at this model over 6 qubits in 1D. To simulate the results one might typically see in a variational quantum algorithm, given that preparing ansa\"tze that can fully represent the target state is often challenging, we approximate an intermediate step of such an algorithm by starting with the exact ground state vector $\ket{\psi_0^{g}}$ for each $g$ and applying random rotations to obtain a ``trial state'' $\ket{\psi_t^{g}}$. Both the direct and moments-based estimates of the staggered magnetisation are then calculated with respect to this trial state after applying various levels of depolarising noise. Figure~\ref{Fig2} shows the interplay between these two phenomena in our simulations for varying average ground state fidelity, $F = \left|\langle \psi_t^{g}  | \psi_0^{g} \rangle\right|^2$, and depolarising noise, ${\rho} \mapsto (1-\frac{3p}{4}){\rho} + \frac{p}{4}(X{\rho}X+Y{\rho}Y+Z{\rho}Z)$ with parameter $p$. We observe that the robustness of the QCM technique to low fidelity trial states and noise, previously only seen in the context of energy problems \cite{hv2020,Vallury2023noiserobustground,mj-chem}, persists when considering arbitrary ground state observables.

We see even for trial states with $\sim$40\% ground state fidelity and 50\% depolarising error, at each value of $g$, the moments based estimate $\langle M \rangle_0^{\rm L(4)}$ provides a powerful correction to the direct measurement of $\langle M \rangle$, remaining remarkably close to the exact value for the ground state staggered magnetisation. This is more or less the level of noise one would expect to see in a typical variational circuit on real hardware with 1\% CNOT error, so the result bodes well for the prospect of applying this technique on present-day devices.

\section{Conclusion}
We have presented an extension of the QCM method to estimate arbitrary ground state observables of quantum systems, with a focus on the Heisenberg model. We observe the robustness of the QCM technique to low fidelity trial states and noise, previously only seen in the context of ground-state energy problems. In particular, the results presented in this paper demonstrate the ability of QCM to estimate ground state properties such as magnetisation and spin-spin correlations of the Heisenberg model to a high precision, even with suboptimal trial states and levels of noise present in near-term quantum hardware. 

Further advancement of this technique will lead to improved understanding of the conditions under which various QCM observable estimates hold greater or lesser validity. We reiterate that the results presented here are from simplified noisy simulations, and the key next steps will involve performing the technique on real quantum hardware and under improved noise models -- previous analysis of the QCM energy estimate in~\cite{Vallury2023noiserobustground} in addition to the preliminary results presented here for the generalisation to arbitrary observables suggests that we should also expect this technique to work quite well on a real noisy device. Furthermore, the examples we have studied here have only scratched the surface of possible Hamiltonians and observables that could be suitable for this approach. Many intriguing problems invite future application of this technique, such as calculating ground state observables for the Fermi-Hubbard model~\cite{stanisic2022observing} and detecting phase transitions in more unconventional systems by measuring nonlocal order parameters~\cite{okada2022identification}.

QCM has so far proven to be an exciting and potentially useful technique for obtaining the ground-state of many body systems on quantum computers, even in the presence of noise and with suboptimal trial states. Our findings here generalise the method to other observables of interest and further highlight its practical potential for solving a wide range of problems on near-term quantum hardware.

\section*{Acknowledgements}
This research was supported by the University of Melbourne through the establishment of the IBM Quantum Network Hub at the University.
HJV is supported by an Australian Government Research Training Program Scholarship.

\printbibliography
\end{document}